\documentclass[%
 reprint,
 floatfix, amsmath,amssymb,
 aps,
]{revtex4-1}

\usepackage{graphicx}% Include figure files
\usepackage{dcolumn}% Align table columns on decimal point
\usepackage{bm}% bold math
\usepackage{pythontex}
\usepackage{amsmath}
\usepackage{hyperref}% add hypertext capabilities
\usepackage[short]{optidef} % optimization problem formatting
\usepackage[version-1-compatibility]{siunitx}
\usepackage[utf8]{inputenc}
\usepackage[usenames,dvipsnames]{xcolor}

\newcommand{\ket}[1]{|#1\rangle}

\begin{document}

\title{Benchmarking Embedded Chain Breaking in Quantum Annealing}

\thanks{This manuscript has been authored by UT-Battelle, LLC, under Contract No.~DE-AC0500OR22725 with the U.S.~Department of Energy. The United States Government retains and the publisher, by accepting the article for publication, acknowledges that the United States Government retains a non-exclusive, paid-up, irrevocable, world-wide license to publish or reproduce the published form of this manuscript, or allow others to do so, for the United States Government purposes. The Department of Energy will provide public access to these results of federally sponsored research in accordance with the DOE Public Access Plan.}

 \author{Erica Grant}
 \email{egrant8@vols.utk.edu}
 \affiliation{%
 Quantum Computing Institute, Oak Ridge National Laboratory, Oak Ridge, Tennessee, 37830, USA
 }
 \affiliation{%
 Bredesen Center for Interdisciplinary Research and Graduate Education, University of Tennessee, Knoxville, Tennessee, 37996, USA
 }

 \author{Travis S.~Humble}%
 \email{humblets@ornl.gov}
 \affiliation{%
 Quantum Computing Institute, Oak Ridge National Laboratory, Oak Ridge, Tennessee, 37830, USA
 }
 \affiliation{%
 Bredesen Center for Interdisciplinary Research and Graduate Education, University of Tennessee, Knoxville, Tennessee, 37996, USA
 }

\begin{abstract}
Quantum annealing solves
%is a novel method for solving a number of challenging real-world computational
combinatorial optimization problems by finding the energetic ground states of an embedded Hamiltonian. However, quantum annealing dynamics 
%that manifests 
under the embedded Hamiltonian 
%there is a potential for quantum annealing to provide an advantage over classical computation for certain problems, there are challenges associated with quantum annealing. In particular, dynamics which
may violate the principles of adiabatic evolution and generate excitations that correspond to errors in the computed solution. Here we empirically benchmark
%investigate the tuning of qubit weights to minimize 
%non-adiabatic dynamics
the probability of chain breaks
%across a range of parameters 
and identify sweet spots for solving a suite of embedded Hamiltonians.
%from quantum annealing in terms of the frequency of chain breaks observed when . 
We further correlate the physical location of chain breaks
%using the and improve the quality of solutions. By pin-pointing where 
in the quantum annealing hardware with the underlying embedding technique and use these localized rates in a tailored post-processing strategies. Our results demonstrate how to use characterization of the quantum annealing hardware to tune the embedded Hamiltonian and remove computational errors.
%for correcting such chain breaks. We demonstrate that post-processing methods using the localized rates of chain breaks offer further improvements for finding the correct solution.
\end{abstract}

\maketitle
%%%%%%%%%%%%%%%%%%%%%%%%%%%%%%%%%%
\section{Introduction}
Quantum annealing (QA) has emerged as a metaheuristic for unconstrained optimization using a combination of quantum and statistical mechanics \cite{farhi2000quantum}. 
QA performs a global search for an optimal solution using quasi-adiabatic dynamics to prepare a distribution over the eigenstate of a time-dependent Hamiltonian that defines a problem of interest. 
Unlike the related paradigm of adiabatic quantum optimization \cite{albash2018adiabatic, AdiabaticQuantumComputingandQuantumAnnealing}, which offers promises for finding the true ground state,
%requires the quantum system to evolve much more slowly than the inverse spectral gap, 
QA  deviates from idealized pure state dynamics predicted by the adiabatic theorem due to non-adiabatic, thermal, and open system effects \cite{chancellor2017modernizing,mishra2018finite,marshall2019power}.
\par 
Recent experimental realizations of QA, i.e., quantum annealers, are neither perfectly adiabatic nor well described by pure quantum states \cite{johnson2011quantum}.  In practice, evolving too quickly violates the adiabatic theorem and introduces a probability for the system to transition to an excited state \cite{farhi2000quantum}, while thermal relaxation and flux noise cause the system to deviate from unitary dynamics \cite{martinis2003decoherence,matsuura2017quantum,novikov2018exploring,marshall2019power}. The additional physics of QA must be accounted when characterizing the  accuracy and performance of this computational method \cite{mcgeoch2013experimental, katzgraber2014glassy, king2015benchmarking, heim2015quantum}.
\par 
As a metaheuristic, strategies for tuning QA have proven useful for demonstrating a variety of domain-specific examples of unconstrained optimization \cite{neven2008image, king2014algorithm, ushijima2017graph, dridi2017prime, venturelli2019reverse}. 
%McGeoch et al. and King et al. have both
Recent demonstrations have used the hardware controls to reduce the effects of bias noise on QA hardware \cite{mcgeoch2013experimental, king2015benchmarking}, while others have tailored the annealing schedule to incorporate non-adiabatic effects to improve upon the probability of success \cite{steiger2015heavy, venturelli2019reverse,marshall2019power,pearson2019analog}. 
However, there are limits to how well the dynamics of quantum annealers can be be controlled and these practical considerations lead to constraints on quantum computational performance \cite{king2014algorithm, king2015benchmarking}. 
\par 
A leading constraint for quantum annealers is the limited connectivity between elements of the quantum register \cite{TS_Humble_2014}. Programming a sparsely connected register requires embedding the logical representation of the input Hamiltonian as a physical representation that accounts for such constraints \cite{choi2011minor,klymko2014adiabatic,boothby2016fast,goodrich2018optimizing,vyskocil2019embedding}. For the embedded Hamiltonian, the computed quantum state is encoded using strongly coupled spins to represent a single logical spin. For the ideal solution state, all the physical spins representing a logical spin perfectly agree in value. A chain break corresponds to a disagreement in the state of the embedded physical spins. Methods to mitigate against chain breaks include post-processing strategies to resolve disagreement in value \cite{pudenz2015quantum,PhysRevA.92.042310,mishra2016performance}. Yet the physics underlying chain breaks make simple post-processing methods fail in certain cases \cite{PhysRevResearch.2.023020}, and insights into the behavior of chains breaks are needed to enable better estimates of both raw and post-processed computational accuracy. 
\par 
Here we investigate the behavior of chain breaks and the impact of embedding on the probability of chain breaks. We evaluate both pre-processing and post-processing strategies that affect the probability of success in finding the true ground state as well as the rate at which chain breaks occur. We use a suite of directly embedded Ising Hamiltonians derived from instances of portfolio optimization, for which the globally optimal states and baseline performance have been reported previously \cite{grant2020benchmarking}. We empirically benchmark the frequency of chain breaks observed as well as the localization within the quantum hardware with respect to problem size and intra-chain coupling. We also demonstrate how post-processing methods using the localized rates of chain breaking improve the probability of success for this benchmark.
\par 
The remainder of the presentation is organized as follows. 
We present a summary of QA for embedded Hamiltonians including a description of embedding and post-processing methods in Sec.~\ref{sec:emb_cs}. In Sec.~\ref{sec:bench}, we present details of the benchmark problems and methods used to study chain breaks. We present results from these benchmarks in Sec.~\ref{sec:results} and final conclusions in Sec.~\ref{sec:conclusions}
%%%%%%%%%%%%%%%%%%%%%%%%%% 
%\section{Quantum Annealing}
%\label{sec:qa}
\section{Quantum Annealing Embedded Hamiltonians \label{sec:emb_cs}} 
In the pure-state representation, quantum annealing solves the problem of finding the ground state of a Hamiltonian $H_1$ by evolving an initial quantum state $\ket{\Psi(0)}$ under the time-dependent Schr\"{o}dinger equation \begin{equation}
\label{Schrodinger Equation}
i \hbar \frac{d}{d t} \ket{\Psi(t)}  = H(t) \ket{\Psi (t)} \hspace{1cm} t \in [0, T]
\end{equation}
with $T$ the total annealing time. Let the idealized time-dependent Hamiltonian be given by
\begin{equation}
\label{Adiabatic Evolution}
H(t) = A(s(t))H_{0} + B(s(t)) H_{1}
\end{equation}
with $s(t) \in [0,1]$ the control schedule and $A(s)$ and $B(s)$ the time-dependent amplitudes satisfying the conditions $A(0)\gg B(0)$ and  $A(1) \ll B(1)$. The initial Hamiltonian $H_0 = - \sum_i^n \sigma_i^x$ sums over the  Pauli-$X$ operators $\sigma_i^x$ representing all $n$ spins. 
\par 
The $j$-th instantaneous eigenstate at time $t$ is defined by
\begin{equation}
\label{instaneous_eigenstates}
H(t) \ket{\Phi_j(t)} = E_j (t) \ket{\Phi_j(t)}
\end{equation}
where $j$ ranges from $0$ to $N-1$ and $N=2^n$ corresponds to the dimension of the Hilbert space. The probability of staying in the ground state throughout the anneal is high as long as the annealing times $T$ is sufficiently long. In particular, $T$ must be much greater than $O({g^{-2}_{min}})$ where $g_{min}$ is the minimum energy gap between the ground state manifold and nearest lying excited states \cite{farhi2000quantum}. 
\par 
We will only consider instances of $H_1$ represented by the (logical) Ising Hamiltonian
\begin{equation}
\begin{aligned}
\label{eq:Ising_Hamiltonian}
H_1 = \sum_{i} h_i \sigma_{i}^z + \sum_{i,j} J_{i,j} \sigma_{i}^z \sigma_{j}^z + \beta
\end{aligned}
\end{equation}
where $h_i$ is the bias on the $i^{th}$ spin, $J_{i,j}$ is the coupling strength between the $i^{th}$ and $j^{th}$ spins, $\sigma_{i}^z$ is the Pauli-$Z$ operator for the $i^{th}$ spin, and $\beta$ is a problem-dependent constant. This form for $H_1$ is well known to express many combinatorial optimization problems \cite{lucas2014ising}.
\par
For our benchmarks, we use the commercial quantum annealers available for research from D-Wave Systems \cite{johnson2011quantum}. Our tests use the D-Wave 2000Q device, which represents a  lattice of up to 2048 programmable superconducting flux qubits. As discussed above, many different sources contribute to non-adiabatic dynamics that cause two types of measurable errors in the logical solution. The first error is an excitation of the system in which the measured sample solution is an excited state which can be caused by Landau-Zener tunneling at level crossings between the lowest energy states or thermal excitations which cause the system to leave ground state \cite{Albash_2015, PhysRevA.65.012322, amin2009decoherence}. The second error appears from thermal or magnetic noise that affects only some physical spins in the chain which causes a broken chain in the sample solution. 
%%%%%%%%%%%%%%%%%%%%%%%%%%%%%%%%%%%%
%\section{Embedding and Chain Strength \label{sec:emb_cs}} 
\subsection{Embedded Hamiltonians}
%Quantum annealers currently have a limited hardware connectivity, for which embedding strategies are used to map highly connected logical problems onto physical representations. Choi first discussed graph minor embedding, aka, embedding, to represent a logical Hamiltonian within a sparsely connected hardware graph \cite{choi2008minor, choi2011minor}. 
\par 
Consider the graphical representation of the logical Hamiltionian $G_l = (V_l, E_l)$ with vertices $V_l$ that represent the $n$ spins and 
%with bias $h_i$  weights and 
edges $E_l$ that represent couplings between these spin sites. 
Embedding maps the logical Hamiltonian $G_l$ to a physical Hamiltonian $G_p$ as $G_l \rightarrow G_p$ by representing each vertex $v_i \in V_l$ by a subgraph, or chain, $T_i$. Edges within each chain $T_i$ specify the intra-chain coupling between physical spins, while edges between different chains specify inter-chain coupling.
The embedded physical Hamiltonian is represented by the graph $G_p = (V_p, E_p)$ with vertices $V_p$ that correspond to physical spin sites and edges $E_p$ that correspond to the inter- and intra-chain couplings between these physical spins.
\par
Embedding selects chains that satisfy the constraints of both the logical graph $G_l$ and  the physical hardware connectivity with graph $G_p$. There are several methods available for embedding graphs into the physical Chimera lattice, which has $8$ qubits per unit cell with a maximum connectivity of $6$ between qubits as seen in Figure~\ref{fig:graphs}. A standard method of embedding is a random embedding algorithm which randomly assigns qubits in chains on the hardware such that all qubits have the necessary connection to satisfy the logical problem. A method for embedding developed by Boothby, King, and Roy based on a clique embedding which is an algorithm which is designed to embed problems with fully connected graphs by minimizing chain length and maximizing the clique which gives more symmetry. Clique embedding typically generates shorter (relative to a random embedding) and uniform chain lengths of size 
\begin{equation}
    l_c = \frac{n}{4} + 1
\label{eq:clique_chain_length}
\end{equation}
for $n$ logical spins \cite{boothby2016fast}. Theory suggests that shorter chain lengths lead to a lower probability of errors caused by noise \cite{Young_2013,venturelli2015quantum, boothby2016fast}. 
\par
Given an embedding with hardware graph $G^{*} = (V^{*}, E^{*})$, the resulting Hamiltonian embedded is specified as
\begin{equation}
    H^* = - \sum_{l \in V^*} h_l^* \sigma^z_l - \sum_{(l,m) \in E^*}J_{l,m}^* \sigma^z_l \sigma^z_m 
\label{eq:embedded_Hamiltonian}
\end{equation}
The physical biases are given by 
\begin{equation}
    h_l^* = \frac{h_i}{|T_i|}
\label{eq:Ising_Coefficient_h}
\end{equation}
for all physical spins $l,m \in V_{T_{i}}$ where $V_{T_{i}}$ is the chain for the $i^{th}$ logical spin, and
\begin{equation}
 J_{l,m}^*= \begin{cases} 
        \frac{J_{i,j}}{\text{edges}(T_i, T_j)} , & \text{ for } l \in T_i, m \in T_j, \text{and } i \neq j\\ 
        k, & \ \text{for } l \in T_i, m \in T_j, \text{and } i = j\\
        0, & \ \text{otherwise}
        \end{cases}
\label{eq:Ising_CoefficientJ}
\end{equation}
where the parameter $k$ represents the intra-chain coupling. The value of $k$ is chosen to ensure intra-chain spins are strongly coupled relative to the magnitude of the inter-chain coupler strength, i.e., $|J_{i,j}|$ which is normalized in be in range $[0, 1]$.
\par 
Ensuring an embedded chain of spins collectively represents a single logical variable requires an intra-chain coupling chain strength $k$ that is larger in magnitude than the inter-chain couplings. In other words, the chain of physical spins must be strongly coupled to remain a single logical spin. However, chains can become ``broken'' in so far as individual physical spins within the chain differ in their final state. 
\par
In general, chain breaks arise from nonadiabatic dynamics that lead to local excitation out of the lowest energy state with theory suggesting that longer chains are more susceptible to these effects \cite{king2014algorithm, Dziarmaga_2005}. King et al.~observed that chains break with higher probability when $k$ is too low, while errors from noise on the hardware such as from non-zero temperature and magnetic interference can be amplified if the magnitude of the intra-chain coupling $|k|$ is much greater than the inter-chain couplings $|J_{i,j}|$. These discrepancies are found to decrease the overall probability of finding the ground state \cite{king2014algorithm}. 
\par 
Venturelli et al.~found that solution quality when solving fully connected graphs on the D-Wave $2$ (an predecessor to the D-Wave 2000Q with up to 1000 physical qubits and the same Chimera structure) varies when tuning $k$ to minimize the number of ground states returned with broken chains \cite{venturelli2015quantum}. Hamerly et al.~ experiments with the D-Wave $2000$Q quantum annealer revealed observations of increased probability to find the ground state (up to $4$ orders of magnitude increase) from increasing chain strength such that the probability of chains breaking reduced to the order of $10^{-1}$ \cite{hamerly2018scaling}. However, there is evidence to suggest that tuning $k$ too far can compress the problem scale and again decrease the probability of success \cite{bian2016mapping}. 
\par 
Previous research on QA performance shows that chain strength plays an important role in quantum annealing performance. Namely, $|k|$ must be sufficient large that the chain continues to represent the logical qubit throughout the anneal. However, if $k$ is too strong the inter-chain couplings can become overpowered which causes probability of success to again decrease \cite{raymond2020improving}. The sweet spot for $k$ partially depends on the inherent noise of the hardware, but how much the optimal value of $k$ depends on problem structure is unknown because the majority of experiments use fully connected graphs.
%%%%%%%%%%%%%%%%%%%%%%%%%%%%%%%%%%%%%%%%%
%\section{Post-processing \label{sec:post}}
\subsection{Post-processing Chains}
An additional control is required for decoding embedded chains that are recovered when measuring the physical spins on the hardware. In the absence of chain breaks, the logical value is inferred directly from the unanimous selection of a single spin state by every physical spin. In the presence of chain breaks, several strategies may be employed to decide the logical value, including majority vote, discard, and a greedy descent \cite{ocean_tools, king2014algorithm}. 
\par 
Majority vote selects the logical spin value as the value that occurs with the highest frequency amongst all spins $q_l$ in a chain. Discard ignores any solutions with broken chains. Greedy descent is a hybrid computing technique that takes the solution with broken chain returned by the D-Wave and feeds it into a classical gradient descent algorithm to locally search for the solution. This greedy descent flips random bits in the broken chains of the solution to find the lowest energy. 
\par 
Reverse annealing can also be used as a post-processing technique for forward annealing. Reverse annealing is a technique can extend a forward anneal by starting from a solution state and annealing in the reverse direction before an optional pause and annealing in the forward direction. King et al.~found small improvements in time to solution by implementing majority vote and an order of magnitude difference when implementing the greedy descent method \cite{king2014algorithm}. Post-processing can be used to interpret results from broken chains. Some of these methods such as discard and majority vote simply attempt to clean up random errors for benchmarking results. However, reverse annealing and greedy descent can be used to apply a local search around the embedded solution returned by the quantum annealer. In the experiments reported below, we consider majority vote and exclude greedy decent and reverse annealing to recover the underlying probability of success from  forward annealing alone.
%%%%%%%%%%%%%%%%%%%%%%%%%%%%%%%%%%%%%%%%%
\section{Benchmarks \label{sec:bench}}
We characterize QA using benchmarks which evaluate the quantum annealers ability to find the lowest energy solution and likelihood for noise to cause errors in the solution. The first benchmark that we use is probability of success 
\begin{equation}
    p_{s} = |\langle\Phi_{0}(T)|\rho|\Phi_{0}(T)\rangle|^2
\label{eq: p_s intro}
\end{equation}
which compares the solution found by the quantum annealer $\rho$ to the global minimum solution $\Phi_{0}(T)$ found with a brute force solver where $\delta_{i} = 1$ if the solution matches the known ground state and $\delta_{i} = 0$ if it does not. For the $o$-th problem instance, the average probability of success is given by
\begin{equation}
\tilde{p}_{s}^{(o)} = \frac{1}{N_s}\sum_{i=1}^{N_s}{\delta_{i}}
\end{equation}
where $N_s$ is the total number of acquired samples. The estimated probability of success for the $o$-th instance is averaged over all problem instances to obtain an average probability of success given by
\begin{equation}
\tilde{p}_{s} = \frac{1}{N_p}\sum_{o}^{N_p}{ \tilde{p}_{s}^{(o)}}.
\end{equation}
where $N_p$ is the total number of problems.
\par
The second metric we use is the number of samples with broken chains observed in each QA solution instance as given by 
\begin{equation}
\begin{aligned}
\tilde{p}_{b}^{(o)} = \frac{1}{N_s}\sum_{i=1}^{N_s}{\epsilon_{i}}
\end{aligned}
\end{equation}
where the statistic $\epsilon_{i} = 1$ when the $i$-th sample solution contains at least one broken chain for any of the logical spins and $\epsilon_{i} = 0$ when no embedded chain is broken. With this metric we are able to observe how chain strength is related to the probability of chains breaking. The average probability of samples with broken chains over many problems is given by
\begin{equation}
\begin{aligned}
\tilde{p}_{b} = \frac{1}{N_p}\sum_{k}^{N_p}{ \tilde{p}_{b}^{(o)}}.
\end{aligned}
\end{equation}
where $N_{p}$ is the number of problem instances.
\par
We also analyze the density of  chain breaks for each problem to determine how chain strength control impact the severity of chain breaks from noise. The average ratio of broken chains per problem is given by
\begin{equation}
\begin{aligned}
\tilde{r}^{(o)}_{b} = \frac{1}{N_s}\sum_{i=1}^{N_s}{ \frac{c_b}{N}}
\end{aligned}
\end{equation}
where $c_b$ is the number of broken chains and $N$ is the number of spins and therefore the total number of chains in the sample. We us this benchmark to plot the average ratio of broken chains for each of the problems for a particular problem size. The average ratio of broken chains for all problems is given by
\begin{equation}
\begin{aligned}
\tilde{r}_{b} =  \frac{1}{N_p}\sum_{k}^{N_p}{\tilde{r}^{(k)}_{b}}.
\end{aligned}
\end{equation}
We use this benchmark to plot the average ratio of problem chain breaks for each problem size. The final benchmark is used to determine the probability for each spin in an intra-coupling to differ from the global minimum solution when a chain breaks $(\tilde{p}_q)$.
\begin{equation}
\begin{aligned}
\tilde{p}_q = \frac{1}{N_b}\sum_{i=1}^{N_b}{ q_b}
\end{aligned}
\end{equation}
where $N_b$ is the number of broken samples for each problem and $q_b$ is a binary variable indicating whether the spin in the broken chain is incorrect. We use this benchmark to plot a heatmap of the probability of each spin to be faulty for all chains in the embedding for each problem size.
%%%%%%%%%%%%%%%%%%%%%%%%%%%%%
%%%%%%%%%%%%%%%%%%%%%%%%%%%%%
\section{Methods}
\label{sec:methods}
We solve portfolio selection using QA to estimate the above metrics. Portfolio selection solves for an optimal percentage of an investor's budget to allocate toward assets under consideration when building a portfolio. We solve a formulation of portfolio selection given a fixed budget $b$, a granularity of percent allocation with factor $w$ where the smallest percentage is $p_w = 1/(2^{w - 1})$, and a historical price data for each asset. The resulting unconstrained optimization problem is \cite{grant2020benchmarking}
\begin{maxi}
  {x}{{\theta_1 \sum^{n}_{i} \sum^{w}_{k = 1} 2^{k - 1} r_i  x_{i}} 
\breakObjective{- \theta_2(\sum^{n}_{i} \sum^{w}_{k = 1} 2^{k - 1}b p_w   x_{i} - b)^2}
\breakObjective{- \theta_3 \sum^{n}_{i, j} \sum^{w}_{k = 1} 2^{k - 1} 2^{k^{\prime} - 1} c_{i, j} x_{i} x_{ j}}}{}{}
{\label{eq:MPO_unconstrained}}
\end{maxi}
where $x_i \in \{0, 1\}$, the problem size is $n = m w$, $c_{i, j} = \frac{p_w^2 \sum^{N_f}_{l = 1}( a_{i,l}-  \bar a_i)( a_{j,l}- \bar a_j)}{N_f -1}$ is the covariance between the historical price data entries for $N_f$ price points between the $i^{th}$ and $j^{th}$ assets, $r_i$ is the expected return for asset $i$, and  $\theta_1, \theta_2$ and $\theta_3$ are Lagrange multipliers used to weight each term for maximization or penalization.
\par
We cast Eq.~(\ref{eq:MPO_unconstrained}) into quadratic unconstrained binary optimization (QUBO) as 
\begin{mini}
{x}{\Big(\sum_{i}^{n} q_{i} x_{i} + \sum_{i, j}^{n} Q_{i, j} x_{i} x_{ j} + \gamma\Big)}{}{}
{\label{eq:QUBO}}
\end{mini}
where $q_i$ is the  $i^{th}$ linear logical bias, $Q_{i,j}$ is the inter-coupler weight representing the interactions between the $i^{th}$ and $j^{th}$ variables, and $\gamma$ is a constant. Note that switching the sign from Eq.~(\ref{eq:MPO_unconstrained}) leads to the QUBO as a minimization problem where the solution represents the optimal portfolio selection. The  relationships between Eq.~(\ref{eq:MPO_unconstrained}) and Eq.~(\ref{eq:QUBO}) are given as
\begin{equation}
\begin{aligned}
& q_i = -\theta_1 r_{i} - 2 \theta_2 b^2 p_w  \\
& Q_{i,j} = \theta_2 b^2 p_{w}^2  +  \theta_3 c_{i,j}  \\
& \gamma = \theta_2 b^2
\end{aligned}
\end{equation}
This QUBO formulation can be easily converted into a classical Ising Hamiltonian 
\begin{equation}
\label{Ising Hamiltonian}
H(s) =  \sum_{i}  s_i h_i + \sum_{i,j}s_i s_j J_{ij} + \beta
\end{equation}
where spin $s_i\in \{-1, 1\}$  with $s = (s_1, s_2, \ldots, s_n)$ is defined by $s_i = 2 x_1 - 1$, $h_i$ is the linear bias, $J_{ij}$ is the coupling, and $\beta$ is a problem-specific constant. The relationship between Eq.~(\ref{eq:QUBO}) and Eq.~(\ref{Ising Hamiltonian}) is given as 
\begin{equation}
\begin{aligned}
& J_{i,j} = \frac{1}{4} Q_{i,j}\\
& h_i = \frac{q_i}{2} + \sum_{j} J_{i,j}\\
& \beta = \frac{1}{4} \sum_{i,j} Q_{i,j} + \frac{1}{2} \sum_{i} q_i + \gamma
\end{aligned}
\end{equation}
The classical Ising formulation is then converted into a corresponding quantum Ising Hamiltonian given by Eq.~\ref{Ising Hamiltonian} using the correspondence $s_i \rightarrow \sigma_{i}^{z}$.
\par
We use a suite of $1000$ problems derived historical price data generated from a uniform random distribution. This data is modeled after stock market volatility by making subsequent prices for each stock a random percent increase or decrease in the range of $\pm 25\%$. We then use this price data with $w=4$, and $b = 1$ to generate each problem using Eq.~\ref{eq:MPO_unconstrained} - \ref{Ising Hamiltonian}. 
\par
\begin{figure}[h!]
\centering
\includegraphics[width=90mm]{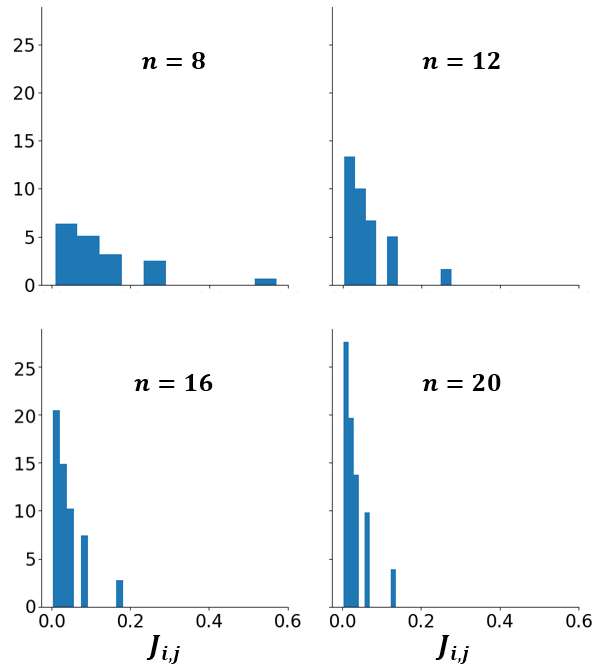}
\caption{A histogram of all $J_{i,j}$ values for $1000$ portfolio optimization problems for each problem size $n$. This graph is normalized to the probability density. }
\label{fig:js}
\end{figure}

 
In Fig.~\ref{fig:js}, we present the histogram of the $J_{i,j}$ coupler values and $h_i$ linear bias for 1000 problems with different problem sizes $n = 8, 12, 16, 20$. Together the $J_{i,j}$ and $k$ values compose the inter-chain and intra-chain coupler weights respectively as shown in Equation~\ref{eq:embedded_Hamiltonian}.
\par
We use the clique embedding described in Section~\ref{sec:emb_cs} for all problem instances. The embedded graphs for the four problems sizes are presented in Fig.~\ref{fig:graphs}. These graphs are embedded onto the D-Wave 2000Q processor which is a programmable quantum annealer with a chimera graph structure.
\par 
All benchmarks listed in Sec.~\ref{sec:bench} are implemented and the $\tilde{p}_s$ is found by comparing the quantum annealing solutions to that of a brute force solver which finds the global minimum solution including any degeneracy. The controls studied include varying the chain strength $k$ to observe the effects on the benchmarks as well as post-processing controls which interpret the raw solution returned from the quantum annealer. Other quantum annealing parameters include anneal time which was set to $100 \mu s$ and spin reversal which was not implemented for these experiments in order to fully observe the effects of noise and chain strength.
 \begin{figure}[h!]
\centering
\includegraphics[width=90mm]{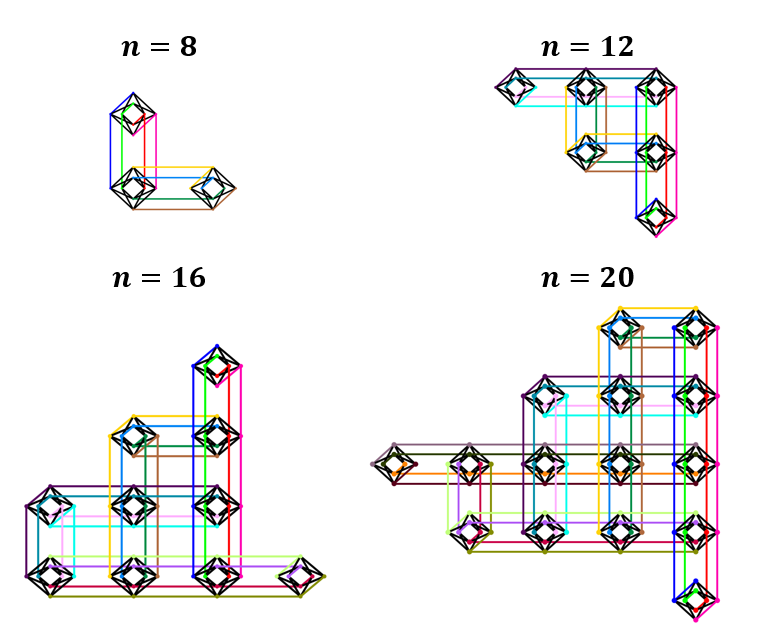}
\caption{The clique embedding graphs with the Chimera graph structure for problem sizes $n$ where each color represents an embedded chain $V_{T_i}$, each diamond grouping of $8$ qubits is a unit cell. }
\label{fig:graphs}
\end{figure}

%After collecting the results from sampling all problems on the D-Wave 2000Q processor, we implement post-processing strategies as mentioned in Section~\ref{sec:emb_cs} to interpret the result of samples with broken chains. 
% \begin{figure}[h!]
% \centering
% \includegraphics[width=85mm]{hs.png}
% \caption{A histogram of all $h_i$ values for $1000$ portfolio optimization problems for each problem size $n$. This graph is normalized to the probability density. }
% \label{fig:hs}
% \end{figure}

\section{Results}
\label{sec:results}
\begin{figure}[h!]
\centering
\includegraphics[width=90mm]{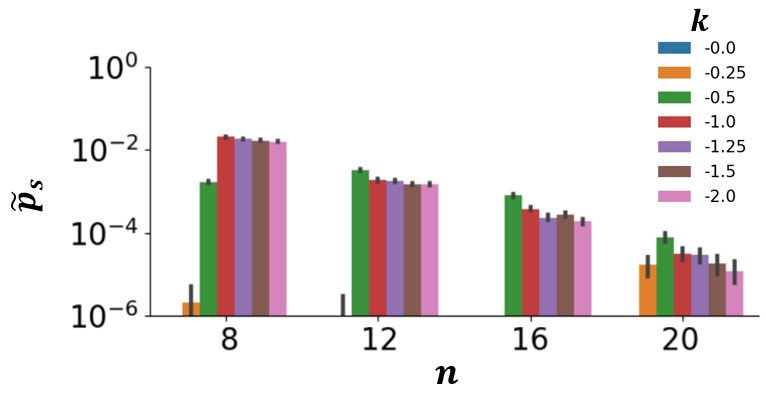}
\caption{The estimated probability of success $\tilde{p}_s$ with respect to intra-chain $k \in [-2,0]$ averaged over $1000$ samples of $1000$ problems for each problem size $n$. All samples with one or more broken chains are discarded (no post-processing) and counted as incorrect.}
\label{fig:discard}
\end{figure}

%% Increasing the chain strength $k$ in Equation~\ref{sec:emb_cs} should reduce the probability of chains breaking, but increasing $k$ too much can result in the intra-coupling strengths $k$ overpowering the inter-chain coupling strengths $J_{i,j}$ and thus increase the probability of the system to jump to an excited state. \ekg{Corrected this sentence to show effects of increasing $k$ too much.} We collect results by sweeping over chain strengths for $k  = [0.0, -0.25, -0.5, -1.0, -1.25, -1.5, -2.0]$ where the strength of $k$ is characterized by how negative the value is \ekg{I put this to be clear that the more negative $k$ is the stronger its weight}. 
%5
%% we are able to observe the $\tilde{p}_s$ as problem size increases for  each $k$. This is visualized in Figure \ref{fig:discard}  where samples with broken chains are discarded and counted as incorrect solutions.  
%%\par 
We first present estimates of the probability of success with respect to chain strength using the discard post-processing method shown in Fig.~\ref{fig:discard}. As expected, $\tilde{p}_s$ vanishes for chains with no intra-chain strength while $\tilde{p}_s$ increases as chain strength $k$ decreases to about $-1.0$. This is followed by a decrease in $\tilde{p}_s$ for smaller $k$, which is an indication that the intra-chain coupling is dominating the spin dynamics. For a fixed value $k$, $\tilde{p}_s$ decreases with increasing problem size $n$ while the optimal $k$ varies with $n$. As shown previously in Fig.~\ref{fig:js}, the distribution of the inter-chain coupling plays an important role in characterizing these problem instances. For the example of $n=8$, the probability of success peaks when the value of $k = -1.0$ for which the maximum $J_{i,j}$ value is $0.57$. However, for all other problem sizes, the optimal intra-chain coupling is observed at $k = -0.5$ for which $J_{i,j}$ is below $0.3$.
\par 
%%%%%%%%%%%%%%%%%%%%%%%%%%%%%%%%%%%%%
%%%%%%%%%%%%%%%%%%%%%%%%%%%%%%%%%%%%%
\par
\begin{figure}[h!]
\centering
\includegraphics[width=90mm]{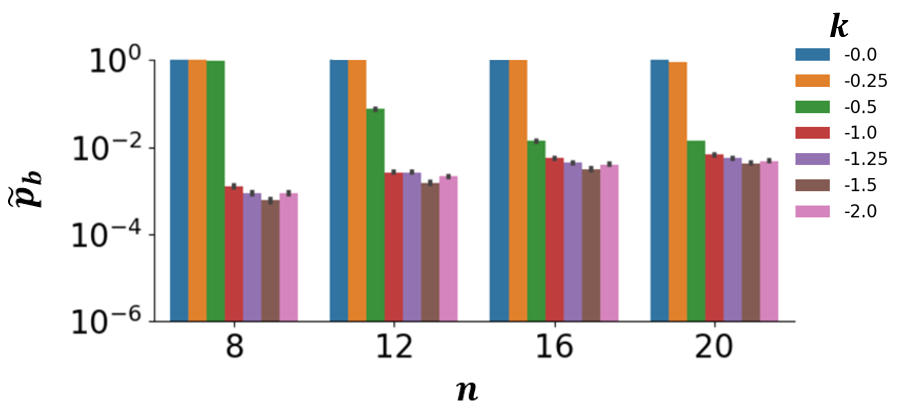}
\caption{The average probability that a sample has at least one broken chain $\tilde{p}_b$ comparing intra-chain strengths $k = [0 \rightarrow -2]$ for $1000$ samples of $1000$ problems for each problem size $n$. }
\label{fig:p_b}
\end{figure}
\begin{figure}[h!]
\centering
\includegraphics[width=90mm]{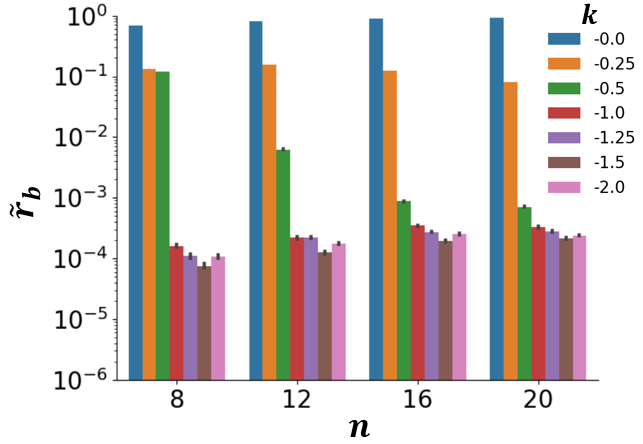}
\caption{The average ratio of broken chains in a sample $\tilde{r}_b$ comparing intra-chain strengths $k = [0 \rightarrow -2]$ for $1000$ samples of $1000$ problems for each problem size $n$.  }
\label{fig:num_breaks}
\end{figure}

We differentiate between the type of errors that occur as $k$ decreases by analyzing the probability of broken chains. In Fig.~\ref{fig:p_b}, we plot the estimated probability of chain breaks $\tilde{p}_b$ with respect to the intra-chain coupling. We observe that samples with $|k| \geq |J_{i,j}|$  have a lower value of $\tilde{p}_b$.  At $k = 0, -0.25,$ and $-0.5$, there is a much higher probability of chains breaking. For weaker $k$, this is shown in Fig.~\ref{fig:num_breaks} where the average ratio of chain breaks $\tilde{r}_b$ reveals that fewer chains break per sample as $k$ increases in strength. At a fixed value of $k = -0.5$, we also observe $\tilde{p}_b$ decreases with increasing $n$. As above, we attribute this behavior to the relative differences in magnitude of the inter-chain coupling $J_{i,j}$ and the increased likelihood for chains to break when $k$ and $J_{i,j}$ are close in value. 
\par 
For all values of $n$, we observe that $\tilde{p}_b$ is significantly smaller for $k > -0.5$ than at $k = -0.5$. Notably, there is a small, but consistent, rise in $\tilde{p}_b$ at $k=-2$ relative to $k = -1.5$. In addition, $\tilde{p}_s$ is reduced in this setting. This combination of behaviors indicates that the errors underlying the reduction in $\tilde{p}_s$ are not from chain breaks but due to sampled solutions from excited states. Consequently, we identity a ``sweet spot'' for the intra-chain coupling that balance the probability of success with probability of chain breaks. The values of $\tilde{p}_b$ and $\tilde{r}_b$ are also higher at $k=-0.5$ than for weaker $k$. More generally, we expect reductions in $\tilde{p}_s$  when $|k|$ is too low in comparison to $|J_{i,j}|$  due to higher $\tilde{p}_b$ and also reduction in $\tilde{p}_s$  when $|k|$ is too high due to solution drawn from the excited states. These results provide a clear link between the intra-chain coupling parameter $k$ and the error rates impacting $\tilde{p}_s$.
%%%%%%%%%%%%%%%%%%%%%%%%%%%%%%%%%%%%%
%%%%%%%%%%%%%%%%%%%%%%%%%%%%%%%%%%%%%
%%%%%%%%%%%%%%%%%%%%%%%%%%%%%%%%%% 
\par
\begin{figure}[h!]
\centering
\includegraphics[width=90mm]{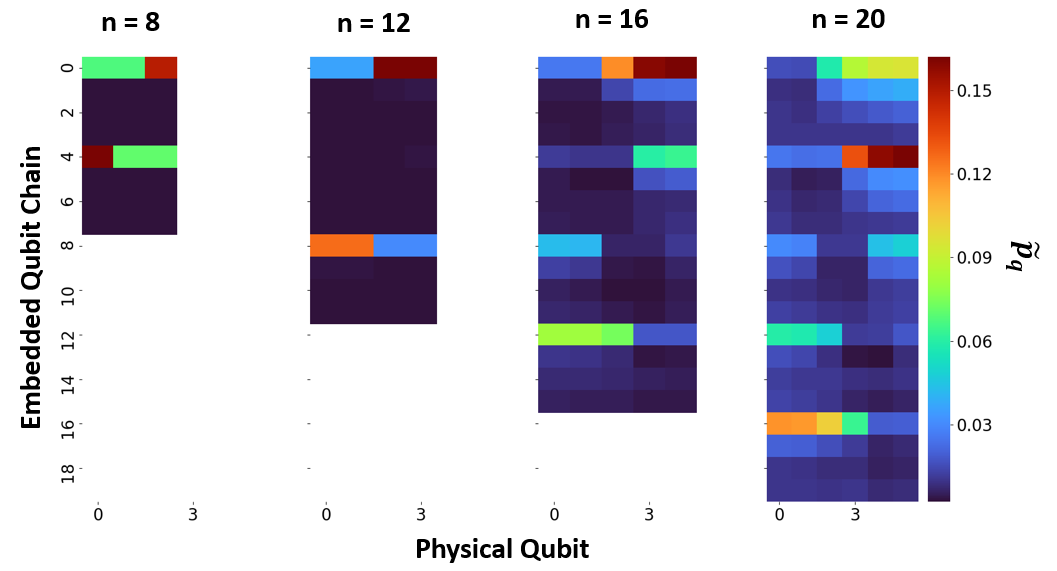}
\caption{A heatmap showing the average probability for each physical spin in a chain to break for a sample over $1000$ problems each with all broken samples where $k = -0.5$. The comparison is between a set of problems from problem sizes $n$.}
\label{fig:breaking_chains_heatmap}
\end{figure}

We further investigate the different types of chains break based on where those breaks occur to identify other factors that play a role in error rates. Figure ~\ref{fig:breaking_chains_heatmap} shows the probability for each spin site in a chain to be faulty when at least on chain in the observed sample was broken. As expected, the probability of a chain breaking increases with chain length, in which the latter correlates directly with the problem size $n$. Notably, breaks that occur with the highest probability are always at the endpoint of the embedded chain.  
\par
%  \begin{figure}[h!]
% \centering
% \includegraphics[width=90mm]{Av_breaks_by_chain.PNG}
% \caption{The average probability for a chain to break for a sample over $1000$ problems each with $1000$ samples where $k = [0 \rightarrow -2]$. The comparison is between a set of problems from problem sizes $n = [8 \rightarrow 20]$.}
% \label{fig:breaking_chains}
% \end{figure}
%%%%%%%%%%%%%%%%%%%%%%%%%%%%%%%%%% 
%%%%%%%%%%%%%%%%%%%%%%%%%%%%%%%%%% 
\par
\begin{figure}[h!]
\centering
\includegraphics[width=90mm]{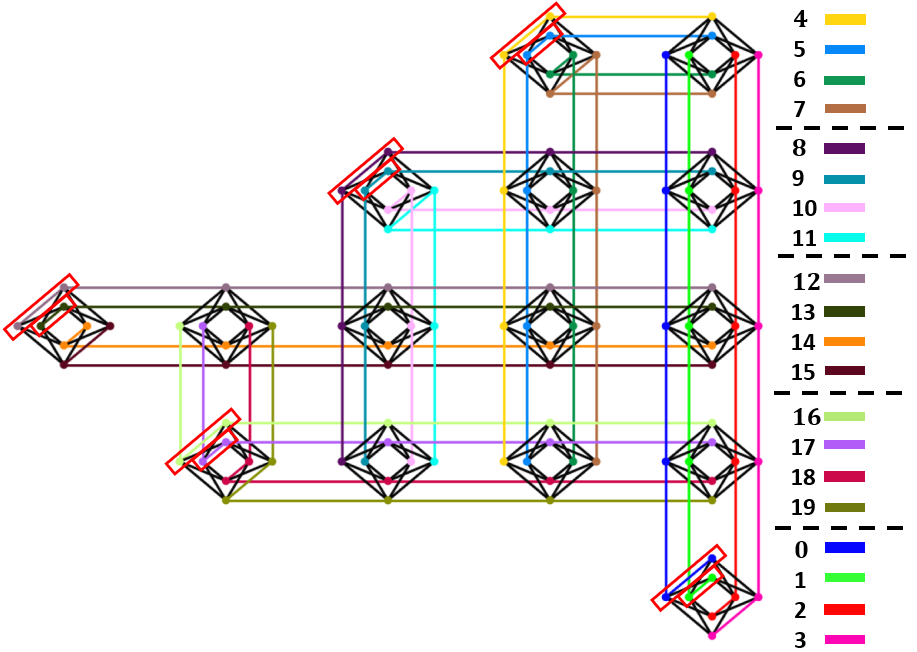}
\caption{The clique embedding graphs for problem size $n = 20$ where the intra-unit cell coupling for embedded chains $0, 4, 8, 12, 16$ are boxed and correspond to the spins which have the lowest probability of being faulty for those chains.}
\label{fig:highlighted_graphs}
\end{figure}

There are additional patterns in the observed chains break that reveal a stronger connection to the embedding. As shown in Fig.~\ref{fig:breaking_chains_heatmap}, there is a higher probability of chain breaks at indices $\{0, 4, 8, 12, 16\}$. The relative variance in the estimated probabilities range $12\%$ to $25\%$, which is sufficient to identify patterns in the chain breaks that correlated with embedding. Comparing Fig.~\ref{fig:breaking_chains_heatmap} with the embeddings shown in Fig.~\ref{fig:graphs} reveals broken chains follow a distinct pattern with clique embedding. As shown in Fig.~\ref{fig:highlighted_graphs}, chains with higher values of $\tilde{p}_q$, that is, at indices 0, 4, 8, 12, 16, represent chains that utilize the top-most physical qubits across all unit cells.
\par 
In addition, the two physical spin sites that break with lowest probability in each of those chains are always coupled within a unit cell as opposed to across unit cells. However, placement on the hardware does not appear to play a strong role in these results. As shown in Fig.~\ref{fig:graphs}, our testing recovered similar behaviors from all four problem sizes while using different location for the embeddings within the hardware lattice. These results indicate that $\tilde{p}_q$ is linked to the hardware embedding and the physical sites in the unit cell but not the specific unit cells that are employed. 
%%%%%%%%%%%%%%%%%%%%%%%%%%%%%%%%%% 
%%%%%%%%%%%%%%%%%%%%%%%%%%%%%%%%%% 
 \begin{figure}[h!]
\centering
\includegraphics[width=90mm]{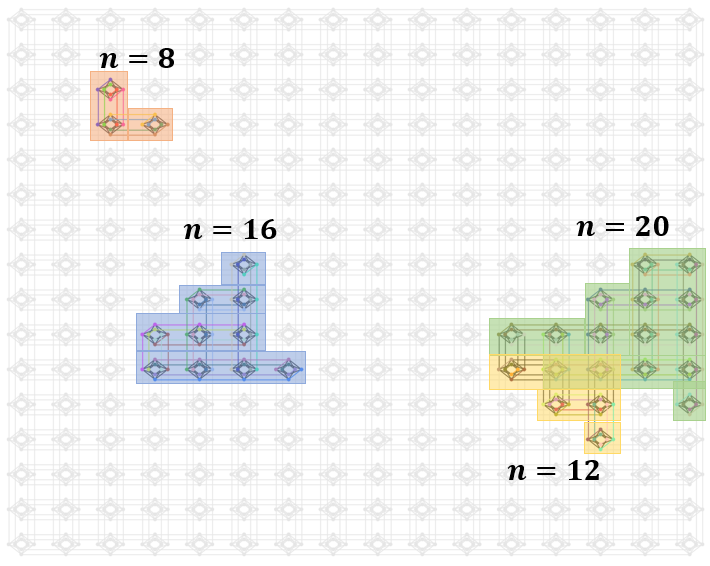}
\caption{The clique embedding graphs used for all 1000 problems on the D-Wave 2000Q hardware with Chimera graph structure for problem sizes $n$.}
\label{fig:graphs}
\end{figure}
%%%%%%%%%%%%%%%%%%%%%%%%%%%%%%%%%% 
\par
We use the detailed information about the frequency and location of chain breaks to tailor methods for post-processing broken chains and assign value to the logical spin. Post-processing strategies that assign a logical value based on the broken chain are more efficient than discarding the sample. We first consider majority vote as a well-known method for post-processing decisions of broken chains. In principle, majority vote works well when the number of errors is small relative to the length of the chain.
%in which the can be a tool to improve $\tilde{p}_s$ for problems where broken chains still retain some representation of the logical qubit. 
However, if the intra-chain coupling $k$ is chosen as too weak, then a majority of sites may be broken, and majority vote performs poorly on average.
\par
\begin{figure}[h!]
\centering
\includegraphics[width=90mm]{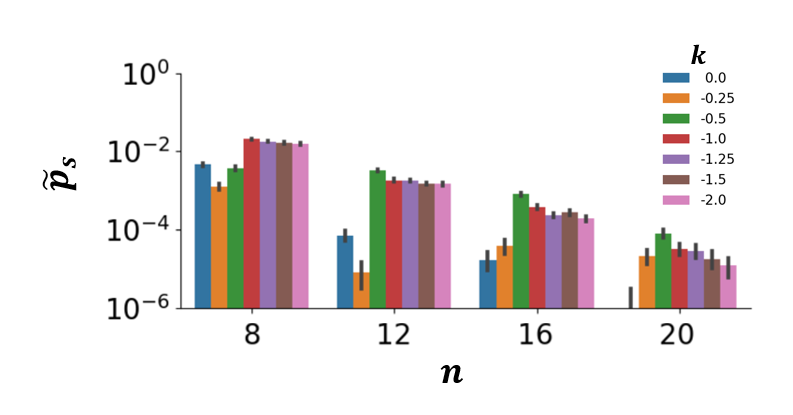}
\caption{The average probability of success over $1000$ problems each with $1000$ samples where $k$ is varied. The comparison is between a set of problems from problem sizes $8$ to $20$. The post-processing method used is majority vote.}
\label{fig:strengths_majority_vote}
\end{figure}

As shown in Fig.~\ref{fig:breaking_chains_heatmap}, we observe that chains which break with high probability have a majority of their sites broken. 
%% For these samples, majority vote is ineffective. 
Post-processing samples from our benchmarks with majority vote yields the probability of success shown in Fig.~\ref{fig:strengths_majority_vote}. Relative to no post-processing, we find that majority vote only increases the probability of success significantly when the probability for chain breaks is greater than 0.1 as is typical for $k > -0.25$. 
%the $\tilde{p}_s$ does not surpass $\tilde{p}_s$ of a random selection $\frac{1}{2^n}$. The $\tilde{p}_b$ is only marginally lower than for $k = 0$, however we observe a $\tilde{p}_s$ that deviates from random selection. %When $n = [8, 12]$, $k = -0.25$ yields $\tilde{p}_s$ an order of magnitude lower than random selection and at $n = 16$ a $\tilde{p}_s$ an order of magnitude higher than random selection. From this, we can conclude that if the chain strength is too low, 
\par 
By contrast, majority vote decreases the probability of success below random selection for $n = 8, 12$ when $k = -0.25$ due to the bias that emerges from chains with many broken sites. 
For $k \geq -0.5$, 
%we see a $\tilde{p}_b$ that is higher than for $k < -0.5$ , but delivers a $\tilde{p}_s$ that is higher than for $k < -0.5$. However, 
majority vote does not improve the probability of success relative to  discarding broken samples. From these observations, we conclude that majority vote does not improve $\tilde{p}_s$ when sufficiently high $|k|$ is present. Indeed, an improvement in probability of success by majority vote can serve as an indicator that the intra-chain coupling is insufficient.
%%%%%%%%%%%%%%%%%%%%%%%%%%%%%%%%%% 
\par
%%%%%%%%%%%%%%%%%%%%%%%%%%%%%%%%%% 
We next apply a post-processing strategy that uses the identified patterns in frequency and location of chain breaks to the benchmark results. Our approach places less weight on sites which are characterized by a higher probability to be faulty within a broken chain. Using the heatmap data shown in Fig.~\ref{fig:breaking_chains_heatmap} for the clique embedding, we assign values for the embedded logical spins by weighting each physical spin value $q_l$ by the probability $\tilde{p}_q$. The resulting decisions are given by 
\begin{equation}
\begin{aligned}
W_i(x_i) = (1 - \prod_{l =0}^{l_c}{(\sigma_l  \tilde{p}^l _q + \sigma^{\prime}_l)} ) \prod_{l =0}^{l_c}{(\sigma^{\prime}_{l}  \tilde{p}^l _q  + \sigma_{l})}
\end{aligned}
\end{equation}
where $W_i(x_i)$ is the score of the value $x_i \in \{-1, 1\}$ for logical spin $i$,  $\sigma_l = (1 + x_i q_l)/2 = 1$ indicates when the value $q_{l}$ of the $l^{th}$ physical spin agrees with the logical choice $x_i$ (and zero otherwise), and $\sigma^{\prime}_l = (1 -x_i q_l)/2 = 1$ indicates when $q_l$ corresponds to the opposing logical choice (and zero otherwise). 
\par
\begin{figure}[h!]
\centering
\includegraphics[width=90mm]{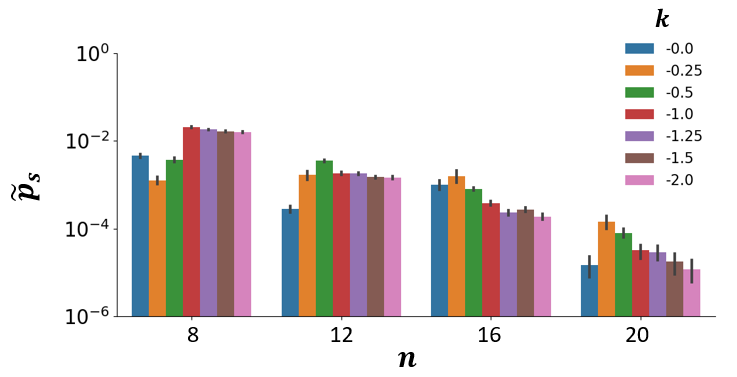}
\caption{The  $\tilde{p}_s$ over $1000$ problems each with $1000$ samples where $k = [0 \rightarrow -2]$. The comparison is between a set of problems from problem size $n$. The post-processing method used is our custom weighted choice technique incorporating the probabilities of faulty chains.}
\label{fig:new_pos}
\end{figure}

 
The results from this weighted voting strategy are shown in Fig.~\ref{fig:new_pos}. We find improvements in $\tilde{p}_s$ for those cases where there is a high $p_b$. This result is far better than a random selection for samples with high $\tilde{p}_b$ which demonstrates that there is still some part of the logical problem which survives when $k$ is too weak.
%%%%%%%%%%%%%%%%%%%%%%%%%%%%%%%%%% 
%%%%%%%%%%%%%%%%%%%
\section{Conclusion}
\label{sec:conclusions}
We have benchmarked the behavior of chain breaks for a suite of embedded Hamiltonians across different sizes and parameter settings. By sweeping over a range of intra-chain strengths, we determined the optimal $k$ with varying problem size $n$. We found that the optimal $k$ is strongly linked to the ratio of $J_{i,j}$ to $k$.
When the coupling $|k|$ is at or below the $|J_{i,j}|$ range in magnitude $|k| \leq |J_{i,j}|$, the probability for chain breaks increases significantly with a corresponding decrease in the probability to find the correct solution. By contrast, when $|k|$ is too high in comparison to $|J_{i,j}|$, then the probability of chain breaks increases only slightly while the probability of success is found to decrease significantly due to excitations. Thus, these benchmarks identify a `sweet spot' for the intra-chain coupling characterized in terms of the metrics for chains breaks and solution success. 
\par 
Our results also characterize the frequency and locations with which chain breaks occur in an experimental quantum annealer. Notably, a pattern emerges in the location with the highest probability of chain breaks for our benchmark suite using the clique embedding. By analyzing the probability for each physical spin in an embedded chain to be faulty, we visualized those locations with the highest probability for faults. We found that there are certain chains that break most frequently as indicated by $\tilde{p}_q$ strongly correlated with the embedding locations. Chains typically break near the chain edges with the spins  coupled within a unit cell statistically least likely to be faulty.  We have applied this information about chain breaks to a tailored post-processing method that demonstrates significant improvements in the probability of finding the correct solution. 
\par
In conclusion, we have shown how benchmarks for chain breaking reveal sweet spots for optimizing probability of success as well as improved techniques for post-processing strategies. While our results have focused on densely coupled instances of portfolio selection using the clique embedding, we expect similar strategies may be applied across other embedded Hamiltonians. These strategies introduce additional options in the design of computational heuristics based on quantum annealing that can be tuned based on hardware characterization. 
%%%%%%%%%%%%%%%%%%%%%%%%%%
\section*{Acknowledgements}
This work was supported by the Department of Energy, Office of Science Early Career Research Program. This research used resources of the Oak Ridge Leadership Computing Facility, which is a DOE Office of Science User Facilities supported by the Oak Ridge National Laboratory under Contract DE-AC05-00OR22725.
\par
We would like to thank Benjamin Stump from Oak Ridge National Laboratory's National Transportation Research Center for aiding in the formulation of equation 22. We would also like to thank Paul Kairys from the Bredesen Center at University of Tennessee for helpful discussion around results in figure 6.

%%%%%%%%%%%%%%%%%%%%%%%%%
%% Bibliography
\bibliographystyle{unsrt}
\bibliography{main.bib}

\end{document}